\documentclass[aps, preprint, groupedaddress, floatfix]{revtex4}

\usepackage{epsfig}
\usepackage{epstopdf}
\usepackage{graphicx}
\usepackage{amsmath}
\usepackage{color}

\begin{document}

\title{Engineering Correlation Effects via Artificially Designed 
Oxide Superlattices}

\author{Hanghui Chen$^{1,2}$, Chris A. Marianetti$^{2}$ and Andrew J. 
Millis$^{1}$}

\affiliation{
 $^1$Department of Physics, Columbia University, New York, NY, 10027, USA\\
 $^2$Department of Applied Physics and Applied Mathematics, Columbia University, New York, NY, 10027, USA\\}
\date{\today}

\begin{abstract}

\textit{Ab initio} 
calculations are used to predict that a superlattice composed of 
layers of LaTiO$_3$ and LaNiO$_3$ alternating along the [001] direction
is a $S=1$ Mott insulator with large magnetic moments on the Ni sites, 
negligible moments on the Ti sites and a charge transfer gap set by the energy 
difference between Ni $d$ and Ti $d$ states, distinct from 
conventional Mott insulators.
Correlation effects are enhanced on the Ni sites via filling the 
oxygen $p$ states and reducing the Ni-O-Ni bond angle. 
Small hole (electron) doping of the superlattice leads to 
a two-dimensional single-band situation with holes (electrons) residing 
on Ni $d_{x^2-y^2}$ (Ti $d_{xy}$) orbital and coupled to antiferromagnetically
correlated spins in the NiO$_2$ layer.
\end{abstract}

\maketitle

Recent advances in atomic scale oxide synthesis provide new approaches
to creating novel electronic and magnetic states in artificially
designed heterostructures~\cite{Hwang-NatMat-2012}. 
Pioneering discoveries include a quasi two-dimensional conducting
electron gas at insulating oxide interfaces~\cite{Hwang-Nature-2002, 
Hwang-Nature-2004, Mannhart-Science-2006, Mannhart-Science-2007, 
Triscone-Nature-2008}, thickness dependent or 
strain-induced metal-insulator transition in ultra thin
films~\cite{Rata-PRL-2008, Yoshimatsu-PRL-2010, Chakhalian-PRL-2011, 
Scherwitzl-PRL-2011}
and at inserted monolayers~\cite{Jang-Science-2012}, as well as
emergent magnetism~\cite{Boris-Science-2011, Brinkman-NatMat-2007, 
Li-NatPhys-2011, Bert-NatPhys-2011}, orbital reconstruction
~\cite{Liu-PRB-2011, Liu-PRL-2012} 
and pseudogap~\cite{Monkman-NatMat-2012}. 
These discoveries mark important milestones towards the goal of using 
the new synthesis capabilities to engineer correlation effects such as 
magnetism and superconductivity. 

In this Letter, we use \textit{ab initio} electronic structure 
methods to demonstrate the viability of internal charge 
transfer and heterostructuring as a route to controlling 
correlation effects in complex 
oxides. The calculations indicate that a superlattice 
composed of layers of LaTiO$_3$ (a narrow gap $S=1/2$ 
antiferromagnet) and LaNiO$_3$ (a wide band paramagnetic metal) 
is a $S=1$ charge transfer Mott insulator, with magnetic moments on 
the Ni sites, negligible magnetism on the Ti sites, and a 
charge transfer gap controlled by the energy difference between 
the Ni and Ti $d$ levels, which is distinct from conventional 
$S=1$ Mott insulators (e.g. NiO and La$_2$NiO$_4$). 
We further show that small hole 
doping may lead to a two-dimensional single-band Fermi surface, 
with carriers residing on Ni $d_{x^2-y^2}$ orbital and strongly 
coupled to correlated spins on Ni $d_{3z^2-r^2}$ orbital, whereas 
doping with electrons also leads to a two-dimensional Fermi 
surface but with carriers on Ti $d_{xy}$ orbital and weakly 
coupled to antiferromagnetic ordering in the NiO$_2$ layer. 

Charge transfer at oxide interfaces has been previously considered, 
for example, in the context of propagation of magnetic order 
from a manganite to a cuprate material~\cite{Chakhalian-Science-2007, 
Chakhalian-NatPhys-2006}, 
or titanate~\cite{Garcia-NatComm-2010},
or nickelate~\cite{Gibert-NatMat-2012, Hoffman-2013, Lee-2013}. 
Here we show that the charge transfer can be 
large enough to fundamentally change the electronic 
properties. In the superlattice considered here, the lone electron in the 
conduction bands of TiO$_2$ layer is completely depopulated 
and the NiO$_2$ layer is transformed from a wide band 
metal to a $S=1$ Mott insulator.

Formal valence changes are also known in related materials, for 
example, $A_2BB'$O$_6$ double perovskites, which can be viewed 
as a [111] direction superlattice~\cite{Serrate-JOP-2007}. 
Experimental indications of 
formal valence changes have been reported for La$_2$MnNiO$_6$
~\cite{Rogado-AdvMater-2005} and La$_2$TiCoO$_6$~\cite{Holmana-JSSC-2007},
but these systems seem not to have been the subject of systematic 
theoretical studies.
Here we provide a comprehensive theoretical 
analysis of this phenomena in the context of a different and more easily 
synthesized [001] superlattice. 
An important feature of the [001] superlattice not found 
in double perovskites is an orbital ordering~\cite{Chaloupka-PRL-2008, 
Hansmann-PRL-2009} with Ni $d_{x^2-y^2}$ at the 
valence band maximum and Ti $d_{xy}$ at the conduction band 
minimum. This orbital ordering leads to a strongly two-dimensional 
character for doped holes and electrons which are coupled to correlated 
spins in the NiO$_2$ layer. 

\begin{figure}[t]
\includegraphics[angle=0,width=12cm]{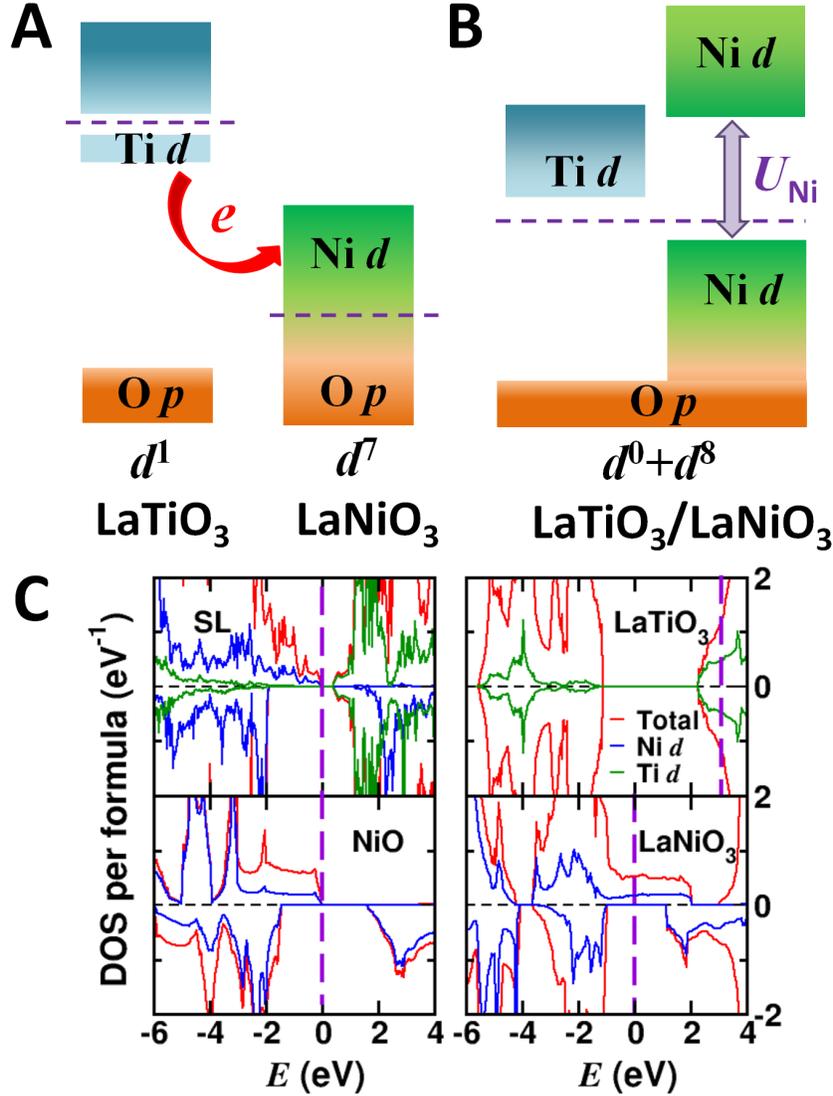}
\caption{\label{fig:charge_transfer} \textbf{A}) Schematic band
  structures of component materials LaTiO$_3$ and LaNiO$_3$. The
  dashed purple lines are the Fermi levels for the two
  materials. LaTiO$_3$ shows insulating behavior with a small
  excitation gap set by Ti $d$-$d$ transitions and a wide energy
  separation between Ti $d$ states and O $p$ states.  LaNiO$_3$
  exhibits metallic behavior with strong mixing between Ni $d$ states
  and O $p$ states. The red arrow highlights the direction of charge
  transfer in the superlattice. \textbf{B}) Schematic band structure
  of (LaTiO$_3$)$_1$/(LaNiO$_3$)$_1$ superlattice. Ti $d$ states are
  above the Fermi level (dashed purple line).  Correlation effects
  split Ni $d$ states into lower and upper Hubbard bands, separated by
  $U_{\textrm{Ni}}$. \textbf{C}) Densities of states for majority (above
  axis) and minority (below axis) spins of superlattice (upper left)
  and reference materials NiO (lower left), LaTiO$_3$ (upper right;
  zero of energy is shifted so that oxygen bands align with those of
  LaNiO$_3$) and LaNiO$_3$ (lower right). The densities of states are 
  obtained using DFT+$U$ calculations with $U_{\textrm{Ni}}$ = 6 eV and
  $U_{\textrm{Ti}}$ = 4 eV.}
\end{figure}

To understand the origin of charge transfer and formal valence changes, 
we sketch the band structure of 
(LaTiO$_3$)$_1$/(LaNiO$_3$)$_1$ superlattice in 
Fig.~\ref{fig:charge_transfer}A.
We begin with bulk LaTiO$_3$, where
the nominal electronic configuration involves one electron in Ti $d$
states ($d^1$), which lie well above the filled oxygen $p$ states.
Bulk LaNiO$_3$, on the other hand, has a nominal electronic configuration 
of Ni $d^7$ with six electrons filling up the $t_{2g}$ shell and one electron in
the $e_g$ orbitals, which overlap in energy with the oxygen
$p$ states. The different electronegativities of Ti and Ni
suggest that in a superlattice, the Ti
$d$ electron may be transferred to the NiO$_2$ plane, leading to a new nominal
electronic configuration: Ti $d^0$ + Ni $d^8$, with strongly enhanced
correlations arising from the half filling of the Ni $e_g$ shell.
 
\begin{figure}[t]
\includegraphics[angle=-90,width=12cm]{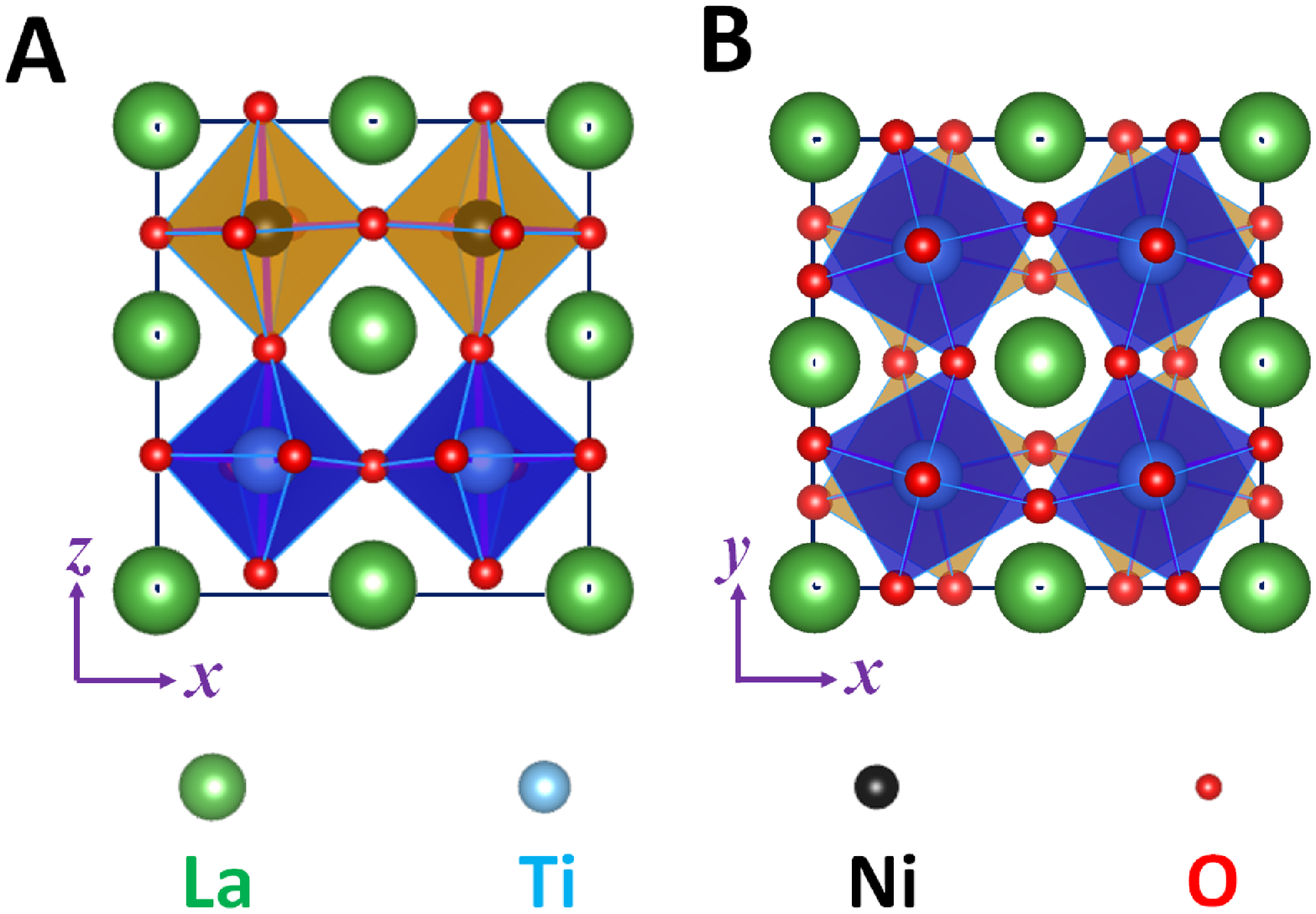}
\caption{\label{fig:atomic_structure} Side view \textbf{A}) and top
  view \textbf{B}) of the unit cell of the
  (LaTiO$_3$)$_1$/(LaNiO$_3$)$_1$ superlattice simulated here. The
  atom positions are obtained from a representative relaxed structure
  found in our first-principles calculations. La atoms are green, Ti
  atoms are blue, Ni atoms are black and O atoms are red. The oxygen
  octahedra shaded orange enclose Ni atoms and the oxygen octahedra
  shaded dark blue enclose Ti atoms.  The side view \textbf{A}) shows
  weak tilting of oxygen octahedra. The top view \textbf{B})
  highlights a strong rotation of oxygen octahedra.}
\end{figure}

To quantitatively test the above idea of engineering correlation effects 
via charge transfer and explore the predicted phenomena,  
we performed density
functional plus $U$ calculations (DFT+$U$) including fully structural 
relaxations
(see Supplementary Material for details). The unit cell used in the
simulations is shown in Fig.~\ref{fig:atomic_structure}. We considered
ferromagnetic ($F$), stripe antiferromagnetic ($S$, wavevector
$(0,\frac{1}{2})$) and checkerboard antiferromagnetic ($G$, wavevector
$(\frac{1}{2},\frac{1}{2})$) states and several values of $U$. The
main conclusions are found to be independent of the nature of the
magnetic order and the value of $U$ (provided that $U$ is within a
physically reasonable range).  The upper left panel of
Fig.~\ref{fig:charge_transfer}C shows the density of states (DOS) of the
superlattice calculated using the physically reasonable
 values $U_{\textrm{Ni}}$ = 6 eV and
$U_{\textrm{Ti}}$ = 4 eV~\cite{Mizokawa-PRB-1995}; 
the ferromagnetic phase is chosen for
clarity of presentation. The bands corresponding to the majority spin
Ni $d$ states are fully occupied, while those corresponding to Ti $d$
states are empty, indicative of a nominal Ti $d^{0}$ + Ni $d^8$
configuration. The superlattice is insulating, with a 0.4 eV energy
gap; the highest occupied states are Ni $d$ states; the lowest
unoccupied states are Ti $d$ states. The energy gap to the Ni
unoccupied $d$ states is much larger, around 1.5 eV.

The other panels of Fig.~\ref{fig:charge_transfer}C show the densities
of states calculated within the same scheme for three reference
materials: face-centered NiO (lower left), cubic perovskite LaTiO$_3$
(upper right, note shift in energy axis) and cubic perovskite
LaNiO$_3$ (lower right). We see that the superlattice DOS strongly
resembles that of NiO, except with Ti $d$ states added in the middle
of the NiO insulating gap.  The densities of states of the two
constituents LaNiO$_3$ and LaTiO$_3$ (right panels of
Fig.~\ref{fig:charge_transfer}C; note the shift in Fermi levels) are
strikingly different. LaNiO$_3$ is metallic, with a large density of
non-Ni (in fact oxygen) states in the vicinity of the Fermi
surface. LaTiO$_3$ has partially occupied Ti $d$ states. The dramatic 
difference in densities of states between the superlattice 
(upper left panel of Fig.~\ref{fig:charge_transfer}C) and the 
constituent materials (right panels of Fig.~\ref{fig:charge_transfer}C) 
shows that an electronic reconstruction has taken place (Note that 
in our calculations LaTiO$_3$ is metallic because we have assumed cubic 
symmetry. A similar calculation but using the experimental structure for 
bulk LaTiO$_3$ produces insulating behavior~\cite{Okatov-EPL-2005}. 
This difference is not important here.).
With O $p$ states of
the two materials aligned (see the right panels of
Fig.~\ref{fig:charge_transfer}C), the lowest occupied Ti $d$ states
are about 3 eV higher than the highest occupied Ni $d$ states, which
is the driving force for the electronic reconstruction, consistent with the
schematics shown in panel A of Fig.~\ref{fig:charge_transfer}. 
We mention here that the actual charge transfer, which we have 
computed from the $z$-dependence of the full 
DFT charge density (see Supplementary Material), is only $\sim 0.2$ electron 
per Ni rather than one electron per Ni as inferred from the formal valence 
changes, because the charge transfer in the near Fermi surface states is to 
a large degree compensated by rehybridization effects similar to those 
discussed in Ref.~\cite{Marianetti-PRL-2004}. 

The electronic reconstruction is also revealed by structural distortions. 
After relaxation of the atomic positions, the apical O atom connecting the
TiO$_2$ and NiO$_2$ layers is found to move about 0.05~\AA~towards the Ti atom,
so that the apical Ni-O bond length is 2.01~\AA~and the Ti-O bond is
1.91~\AA, as expected if charge is transferred from Ti to Ni.
The in-plane Ti-O and Ni-O bond lengths are, on the other
hand, much closer to each other and to the average of the apical bond
lengths: 1.96~\AA~versus 1.94~\AA. We therefore have an
unusual situation of a large Jahn-Teller-type distortion about nominally
spherical Ni ($S$=1) and Ti ($S$=0) ions. In addition to the bond lengths,
the bond angles are also different from bulk values. In the superlattice,
the Ni-O-Ni bond angle is found to be 157$^{\circ}$, 
compared to the bulk values of 
165$^{\circ}$ (experiment~\cite{Torrance-PRB-1992}) and 
168$^{\circ}$ (theory~\cite{Gou-PRB-2011}). The decreased Ni-O-Ni 
bond angle reduces the in-plane inter-Ni hopping and helps stabilize 
the Mott insulating state.  
Both the bond lengths and bond angles have a weak
dependence on $U$; the numbers cited here are obtained from
calculations at $U_{\textrm{Ni}}$ = 6 eV and $U_{\textrm{Ti}}$ = 4 eV,
but the 0.1~\AA~difference between the apical Ni-O and Ti-O bond
lengths and a significantly decreased Ni-O-Ni bond angle are found 
at all the $U$ values studied.

\begin{figure}[t]
\includegraphics[angle=-90,width=12cm]{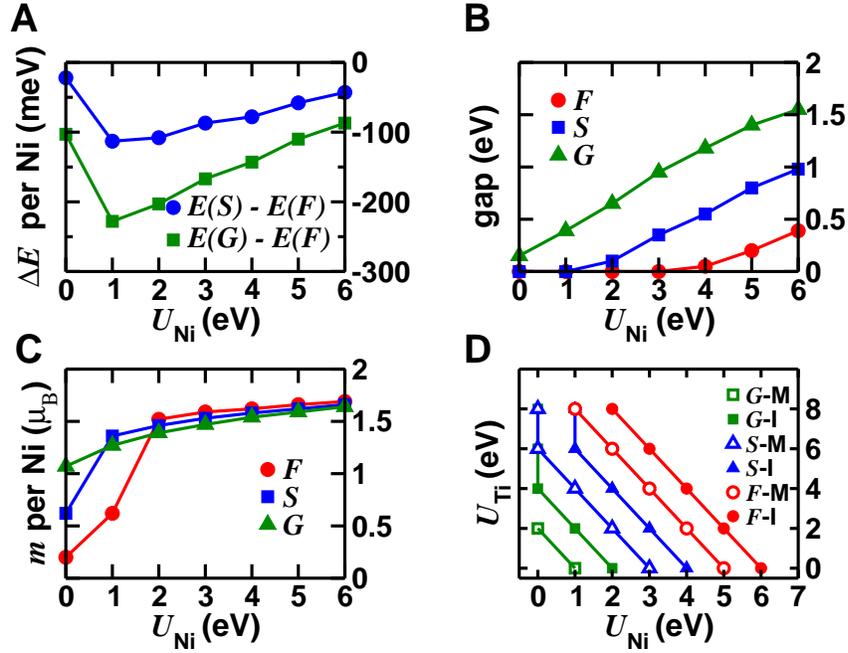}
\caption{\label{fig:electronic_structure} Panels
  \textbf{A}--\textbf{C}: electronic and magnetic properties of
  (LaTiO$_3$)$_1$/(LaNiO$_3$)$_1$ superlattice for different magnetic
  orderings ($F$ denotes ferromagnetic ordering, $S$ denotes $S$-type
  or stripe antiferromagnetic ordering and $G$ denotes $G$-type or
  checkerboard-pattern antiferromagnetic ordering) as a function of
  $U_{\mathrm{Ni}}$ with $U_{\textrm{Ti}}$ = 4 eV. \textbf{A}) Energy
  difference per Ni between different magnetic orderings. \textbf{B})
  Minimum excitation gap. \textbf{C}) Absolute value of site projected
  magnetization of Ni $d$ states. \textbf{D}) Metal-insulator
  boundaries for each magnetic ordering on the $(U_{\textrm{Ni}}, U_{\textrm{Ti}})$ 
  phase diagram, respectively. $F$, $S$ and $G$ have the same meaning as in
  \textbf{A})--\textbf{C}). `M' and `I' denote metallic and insulating
  states. The open symbols denote the upper limits of
  $U_{\textrm{Ni}}$ for which metallic phases are found, while the
  solid symbols indicate the lower limits of $U_{\textrm{Ni}}$ for
  which insulating phases have been found (note that only integer
  values of $U$ have been studied).}
\end{figure}

Within the DFT+$U$ approximation for a wide range of
$U_{\textrm{Ni}}$, we have studied three locally stable states:
ferromagnetic ($F$), stripe ($S$) antiferromagnetic ordering (in-plane
wave-vector $(0,\frac{1}{2})$) and checkerboard ($G$)
antiferromagnetic ordering (in-plane wave-vector
$(\frac{1}{2},\frac{1}{2})$). Panel A of
Fig.~\ref{fig:electronic_structure} shows the energy differences
between these three states. We see that at all values of
$U_{\textrm{Ni}}$ the $G$-type antiferromagnetism has the lowest
energy. As $U_{\textrm{Ni}}$ becomes greater than 1 eV, the energy
differences decrease, consistent with the notion that the NiO$_2$
planes are in a Mott insulating state for which the magnetic energies
scale as $J\sim t^2/U$. 

\begin{figure}[t]
\includegraphics[angle=-90,width=12cm]{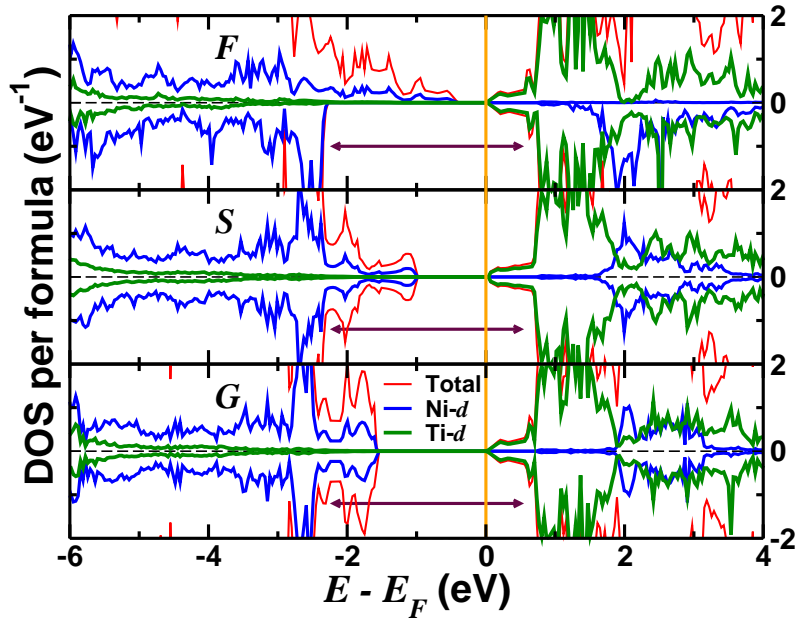}
\caption{\label{fig:mag_compare} Comparison of densities of states for
  ferromagnetic ($F$), stripe antiferromagnetic ($S$) and checkerboard
  ($G$) magnetic orderings calculated at $U_{\textrm{Ni}}$ = 6 eV and
  $U_{\textrm{Ti}}$ = 4 eV.  The conduction band minimum is aligned
  (highlighted by the orange line). The energy gap between the
  conduction band minimum and Ni $d$ main peak (denoted by the maroon
  arrow) is approximately the same for different magnetic orderings.}
\end{figure}

Panel B of Fig.~\ref{fig:electronic_structure} shows the fundamental energy gap
of the superlattice (which, as seen from the upper left panel of
Fig.~\ref{fig:charge_transfer}C, is the energy difference between
Ni $d$ states and Ti $d$ states). We see that in all cases, the
fundamental gap increases linearly with $U_{\textrm{Ni}}$ and with the
same slope, indicating that the nature of the magnetic order only
affects the onset of insulating behavior and not the basic properties
of the state, again consistent with the idea that a Mott insulator has
been created. Further insight into this phenomenon comes from the
densities of states shown in Fig.~\ref{fig:mag_compare}. We see that
changing the nature of the magnetic ordering modifies the details of
the highest occupied states, but does not affect basic features such
as the energy splitting between the Ni $d$ main peak (Ni $t_{2g}$
states) and Ti $d$ states (conduction band minimum), as highlighted by
the maroon arrow in Fig.~\ref{fig:mag_compare}. Panel C of
Fig.~\ref{fig:electronic_structure} shows the magnetic moment on the
Ni $d$ site (computed using the VASP default atomic projector for
Ni--see Supplementary Material). We see that for $U_{\textrm{Ni}} >$ 2 eV,
the magnitude of the local magnetic moment is essentially independent
of the nature of the ordering, is weakly $U$-dependent and is close to the
value expected for the naive atomic $S=1$ states, once again
confirming the Mott nature of the state predicted here.

To this point, our calculations have used the physically accepted
value $U_{\textrm{Ti}}$ = 4 eV~\cite{Mizokawa-PRB-1995}. We have
investigated the robustness of our results to the choice of $U$ by
sweeping the phase space spanned by $U_{\textrm{Ti}}$ and
$U_{\textrm{Ni}}$ for the $F$, $S$ and $G$ magnetic states.  We find
the energy sequence $G < S < F$ throughout the phase space. Figure
~\ref{fig:electronic_structure}D shows that although the position of
the metal-insulator transition boundaries depends on the $U$ values,
the essential features are $U$-independent. The phase boundary for all
the magnetic states has a negative slope, which is understood as
follows: within DFT+$U$, $U_{\textrm{Ti}}$ increases the Ti $t_{2g}$
states because Ti $t_{2g}$ shell is less than half filled, while
$U_{\textrm{Ni}}$ increases the energy gap between the Ni $e_{g}$
majority and minority spins. Therefore, with a given
$U_{\textrm{Ti}}$, we need a $U_{\textrm{Ni}}$ large enough to
separate Ni $e_{g}$ and Ti $t_{2g}$ states and open a fundamental gap
and vice versa.  With $U_{\textrm{Ni}}$ 6 eV or larger, Mott physics
dominates, as the system is rendered insulating for all the magnetic
orderings and all the values of $U_{\textrm{Ti}}$.

\begin{figure}[t]
\includegraphics[angle=-90,width=12cm]{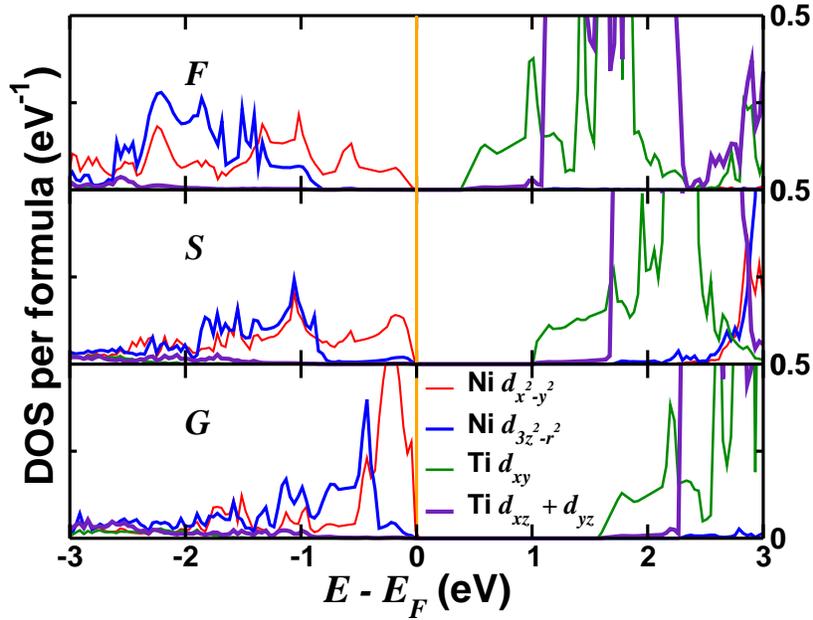}
\caption{\label{fig:orbital-selective} Orbitally resolved Ni $e_g$
  and Ti $t_{2g}$ densities of states (red: Ni $d_{x^2-y^2}$; blue: 
  Ni $d_{3z^2-r^2}$; green: Ti $d_{xy}$; purple: Ti $d_{xz}+d_{yz}$), 
  computed at
  $U_{\textrm{Ni}}$ = 6 eV and $U_{\textrm{Ti}}$ = 4 eV for
  ferromagnetic ($F$), stripe antiferromagnetic ($S$) and checkerboard
  antiferromagnetic ($G$) orderings. }
\end{figure}

The most important experimental test of our calculations is that the
superlattice should be a quasi two-dimensional $S=1$ magnetic
insulator, with the moments residing on the Ni sites and the
correlations most likely antiferromagnetic. The position of the apical
oxygen (nearer to the Ti than the Ni) and the reduced Ni-O-Ni bond angle are 
other important indicators of the predicted electronic reconstruction.  
Orbital splitting is also a test of our calculations.
Fig.~\ref{fig:orbital-selective} presents
the orbitally resolved densities of states for a representative case
($U_{\textrm{Ni}}$ = 6 eV and $U_{\textrm{Ti}}$ = 4 eV). In the
majority spin channel both Ni $d_{x^2-y^2}$ and $d_{3z^2-r^2}$
orbitals are occupied, leading to a spin $S=1$ configuration, while 
all Ti $t_{2g}$ orbitals are empty, resulting in a spin $S=0$ configuration. 
Since both Ni $e_g$ orbitals are filled and all the three Ti $t_{2g}$ 
orbitals are empty, there is no orbital polarization in
the sense of Ref.~\cite{Benckiser-NatMater-2011, Han-PRL-2011}. 
However, our first-principles calculations show that the valence 
band maximum is mainly of Ni $d_{x^2-y^2}$ character and the conduction 
band minimum is dominantly of Ti $d_{xy}$ character.
This orbital ordering indicates that small hole doping may lead to
an orbitally selective Mott metallic state~\cite{Anisimov-EPJB-2002, 
DeMedici-PRL-2009} with a single orbital
(Ni $d_{x^2-y^2}$) active at the Fermi surface but coupled to
antiferromagnetically correlated spins on the Ni $d_{3z^2-r^2}$ orbitals.
In this case, the carriers' motion is strongly affected by magnetism, 
which is similar to La$_{2-x}$Sr$_x$NiO$_4$.
However, on the other hand, small electron doping may result in a 
Ti $d_{xy}$ single orbital Fermi surface, in which the carriers 
are only weakly coupled to the antiferromagnetic ordering in the 
NiO$_2$ layer. This feature is absent in conventional doped Mott 
insulators.
The comparison of carriers' strong/weak coupling to 
correlated spins in one system poses an interesting open question and 
remains to be explored in experiment. Resonant x-ray
photoemission may shed light on the fundamental differences between 
these predicted phases.

In conclusion, we have shown within the DFT+$U$ approximation that 
in a (LaTiO$_3$)$_1$/(LaNiO$_3$)$_1$ superlattice, correlation 
effects are greatly enhanced on the Ni sites, producing
a $S=1$ Mott insulator with an unusual charge transfer 
gap set by Ni $d$ and Ti $d$ states, distinct from conventional 
Mott insulators. The superlattice structure leads to a single 
orbital character at band edges and indicates interesting new 
physics with carrier doping, which is absent in double perovskites.
We believe that DFT+$U$ approximation provides a reliable description 
of correlation effects for stoichiometric and ordered materials. 
More sophisticated methods (e.g. DFT + dynamical mean field theory) 
might be needed to 
treat doped systems with mobile carriers. 
Our findings open a new direction for the control of correlation 
effects via charge transfer and heterostructuring in perovskite 
oxides.

We thank Charles Ahn and Sohrab Ismail-Beigi for
helpful conversations. This research was supported by the Army
Research Office under ARO-Ph 56032 and the United States Department of
Energy under grant DOE-ER-046169.


\begin{thebibliography}{38}
\expandafter\ifx\csname natexlab\endcsname\relax\def\natexlab#1{#1}\fi
\expandafter\ifx\csname bibnamefont\endcsname\relax
  \def\bibnamefont#1{#1}\fi
\expandafter\ifx\csname bibfnamefont\endcsname\relax
  \def\bibfnamefont#1{#1}\fi
\expandafter\ifx\csname citenamefont\endcsname\relax
  \def\citenamefont#1{#1}\fi
\expandafter\ifx\csname url\endcsname\relax
  \def\url#1{\texttt{#1}}\fi
\expandafter\ifx\csname urlprefix\endcsname\relax\def\urlprefix{URL }\fi
\providecommand{\bibinfo}[2]{#2}
\providecommand{\eprint}[2][]{\url{#2}}

\bibitem[{\citenamefont{Hwang et~al.}(2012)\citenamefont{Hwang, Iwasa,
  Kawasaki, Keimer, Nagaosa, and Tokura}}]{Hwang-NatMat-2012}
\bibinfo{author}{\bibfnamefont{H.}~\bibnamefont{Hwang}},
  \bibinfo{author}{\bibfnamefont{Y.}~\bibnamefont{Iwasa}},
  \bibinfo{author}{\bibfnamefont{M.}~\bibnamefont{Kawasaki}},
  \bibinfo{author}{\bibfnamefont{B.}~\bibnamefont{Keimer}},
  \bibinfo{author}{\bibfnamefont{N.}~\bibnamefont{Nagaosa}}, \bibnamefont{and}
  \bibinfo{author}{\bibfnamefont{Y.}~\bibnamefont{Tokura}},
  \bibinfo{journal}{Nature Mater.} \textbf{\bibinfo{volume}{11}},
  \bibinfo{pages}{103} (\bibinfo{year}{2012}).

\bibitem[{\citenamefont{Ohtomo et~al.}(2004)\citenamefont{Ohtomo, Muller,
  Grazul, and Hwang}}]{Hwang-Nature-2002}
\bibinfo{author}{\bibfnamefont{A.}~\bibnamefont{Ohtomo}},
  \bibinfo{author}{\bibfnamefont{D.~A.} \bibnamefont{Muller}},
  \bibinfo{author}{\bibfnamefont{J.~L.} \bibnamefont{Grazul}},
  \bibnamefont{and} \bibinfo{author}{\bibfnamefont{H.~Y.} \bibnamefont{Hwang}},
  \bibinfo{journal}{Nature} \textbf{\bibinfo{volume}{427}},
  \bibinfo{pages}{423} (\bibinfo{year}{2004}).

\bibitem[{\citenamefont{Ohtomo and Hwang}(2004)}]{Hwang-Nature-2004}
\bibinfo{author}{\bibfnamefont{A.}~\bibnamefont{Ohtomo}} \bibnamefont{and}
  \bibinfo{author}{\bibfnamefont{H.~Y.} \bibnamefont{Hwang}},
  \bibinfo{journal}{Nature} \textbf{\bibinfo{volume}{427}},
  \bibinfo{pages}{423} (\bibinfo{year}{2004}).

\bibitem[{\citenamefont{Thiel et~al.}(2006)\citenamefont{Thiel, Hammerl,
  Schmehl, Schneider, and Mannhart}}]{Mannhart-Science-2006}
\bibinfo{author}{\bibfnamefont{S.}~\bibnamefont{Thiel}},
  \bibinfo{author}{\bibfnamefont{G.}~\bibnamefont{Hammerl}},
  \bibinfo{author}{\bibfnamefont{A.}~\bibnamefont{Schmehl}},
  \bibinfo{author}{\bibfnamefont{C.~W.} \bibnamefont{Schneider}},
  \bibnamefont{and} \bibinfo{author}{\bibfnamefont{J.}~\bibnamefont{Mannhart}},
  \bibinfo{journal}{Science} \textbf{\bibinfo{volume}{313}},
  \bibinfo{pages}{1942} (\bibinfo{year}{2006}).

\bibitem[{\citenamefont{Reyren et~al.}(2007)\citenamefont{Reyren, Thiel,
  Caviglia, Kourkoutis, Hammerl, Ricther, Schneider, Kopp, R\"{u}etschi,
  Jaccard et~al.}}]{Mannhart-Science-2007}
\bibinfo{author}{\bibfnamefont{N.}~\bibnamefont{Reyren}},
  \bibinfo{author}{\bibfnamefont{S.}~\bibnamefont{Thiel}},
  \bibinfo{author}{\bibfnamefont{A.~D.} \bibnamefont{Caviglia}},
  \bibinfo{author}{\bibfnamefont{L.~F.} \bibnamefont{Kourkoutis}},
  \bibinfo{author}{\bibfnamefont{G.}~\bibnamefont{Hammerl}},
  \bibinfo{author}{\bibfnamefont{C.}~\bibnamefont{Ricther}},
  \bibinfo{author}{\bibfnamefont{C.~W.} \bibnamefont{Schneider}},
  \bibinfo{author}{\bibfnamefont{T.}~\bibnamefont{Kopp}},
  \bibinfo{author}{\bibfnamefont{A.~S.} \bibnamefont{R\"{u}etschi}},
  \bibinfo{author}{\bibfnamefont{D.}~\bibnamefont{Jaccard}},
  \bibnamefont{et~al.}, \bibinfo{journal}{Science}
  \textbf{\bibinfo{volume}{317}}, \bibinfo{pages}{1196} (\bibinfo{year}{2007}).

\bibitem[{\citenamefont{Caviglia et~al.}(2008)\citenamefont{Caviglia, Gariglio,
  Reyren, Jaccard, Schneider, Gabay, Thiel, Hammerl, Mannhart, and
  Triscone}}]{Triscone-Nature-2008}
\bibinfo{author}{\bibfnamefont{A.~D.} \bibnamefont{Caviglia}},
  \bibinfo{author}{\bibfnamefont{S.}~\bibnamefont{Gariglio}},
  \bibinfo{author}{\bibfnamefont{N.}~\bibnamefont{Reyren}},
  \bibinfo{author}{\bibfnamefont{D.}~\bibnamefont{Jaccard}},
  \bibinfo{author}{\bibfnamefont{T.}~\bibnamefont{Schneider}},
  \bibinfo{author}{\bibfnamefont{M.}~\bibnamefont{Gabay}},
  \bibinfo{author}{\bibfnamefont{S.}~\bibnamefont{Thiel}},
  \bibinfo{author}{\bibfnamefont{G.}~\bibnamefont{Hammerl}},
  \bibinfo{author}{\bibfnamefont{J.}~\bibnamefont{Mannhart}}, \bibnamefont{and}
  \bibinfo{author}{\bibfnamefont{J.~M.} \bibnamefont{Triscone}},
  \bibinfo{journal}{Nature} \textbf{\bibinfo{volume}{456}},
  \bibinfo{pages}{624} (\bibinfo{year}{2008}).

\bibitem[{\citenamefont{Rata et~al.}(2008)\citenamefont{Rata, Herklotz, Nenkov,
  Schultz, and D\"orr}}]{Rata-PRL-2008}
\bibinfo{author}{\bibfnamefont{A.~D.} \bibnamefont{Rata}},
  \bibinfo{author}{\bibfnamefont{A.}~\bibnamefont{Herklotz}},
  \bibinfo{author}{\bibfnamefont{K.}~\bibnamefont{Nenkov}},
  \bibinfo{author}{\bibfnamefont{L.}~\bibnamefont{Schultz}}, \bibnamefont{and}
  \bibinfo{author}{\bibfnamefont{K.}~\bibnamefont{D\"orr}},
  \bibinfo{journal}{Phys. Rev. Lett.} \textbf{\bibinfo{volume}{100}},
  \bibinfo{pages}{076401} (\bibinfo{year}{2008}).

\bibitem[{\citenamefont{Yoshimatsu et~al.}(2010)\citenamefont{Yoshimatsu,
  Okabe, Kumigashira, Okamoto, Aizaki, Fujimori, and
  Oshima}}]{Yoshimatsu-PRL-2010}
\bibinfo{author}{\bibfnamefont{K.}~\bibnamefont{Yoshimatsu}},
  \bibinfo{author}{\bibfnamefont{T.}~\bibnamefont{Okabe}},
  \bibinfo{author}{\bibfnamefont{H.}~\bibnamefont{Kumigashira}},
  \bibinfo{author}{\bibfnamefont{S.}~\bibnamefont{Okamoto}},
  \bibinfo{author}{\bibfnamefont{S.}~\bibnamefont{Aizaki}},
  \bibinfo{author}{\bibfnamefont{A.}~\bibnamefont{Fujimori}}, \bibnamefont{and}
  \bibinfo{author}{\bibfnamefont{M.}~\bibnamefont{Oshima}},
  \bibinfo{journal}{Phys. Rev. Lett.} \textbf{\bibinfo{volume}{104}},
  \bibinfo{pages}{147601} (\bibinfo{year}{2010}).

\bibitem[{\citenamefont{Chakhalian et~al.}(2011)\citenamefont{Chakhalian,
  Rondinelli, Liu, Gray, Kareev, Moon, Prasai, Cohn, Varela, Tung
  et~al.}}]{Chakhalian-PRL-2011}
\bibinfo{author}{\bibfnamefont{J.}~\bibnamefont{Chakhalian}},
  \bibinfo{author}{\bibfnamefont{J.~M.} \bibnamefont{Rondinelli}},
  \bibinfo{author}{\bibfnamefont{J.}~\bibnamefont{Liu}},
  \bibinfo{author}{\bibfnamefont{B.~A.} \bibnamefont{Gray}},
  \bibinfo{author}{\bibfnamefont{M.}~\bibnamefont{Kareev}},
  \bibinfo{author}{\bibfnamefont{E.~J.} \bibnamefont{Moon}},
  \bibinfo{author}{\bibfnamefont{N.}~\bibnamefont{Prasai}},
  \bibinfo{author}{\bibfnamefont{J.~L.} \bibnamefont{Cohn}},
  \bibinfo{author}{\bibfnamefont{M.}~\bibnamefont{Varela}},
  \bibinfo{author}{\bibfnamefont{I.~C.} \bibnamefont{Tung}},
  \bibnamefont{et~al.}, \bibinfo{journal}{Phys. Rev. Lett.}
  \textbf{\bibinfo{volume}{107}}, \bibinfo{pages}{116805}
  (\bibinfo{year}{2011}).

\bibitem[{\citenamefont{Scherwitzl et~al.}(2011)\citenamefont{Scherwitzl,
  Gariglio, Gabay, Zubko, Gibert, and Triscone}}]{Scherwitzl-PRL-2011}
\bibinfo{author}{\bibfnamefont{R.}~\bibnamefont{Scherwitzl}},
  \bibinfo{author}{\bibfnamefont{S.}~\bibnamefont{Gariglio}},
  \bibinfo{author}{\bibfnamefont{M.}~\bibnamefont{Gabay}},
  \bibinfo{author}{\bibfnamefont{P.}~\bibnamefont{Zubko}},
  \bibinfo{author}{\bibfnamefont{M.}~\bibnamefont{Gibert}}, \bibnamefont{and}
  \bibinfo{author}{\bibfnamefont{J.-M.} \bibnamefont{Triscone}},
  \bibinfo{journal}{Phys. Rev. Lett.} \textbf{\bibinfo{volume}{106}},
  \bibinfo{pages}{246403} (\bibinfo{year}{2011}).

\bibitem[{\citenamefont{Jang et~al.}(2011)\citenamefont{Jang, Felker, Bark,
  Wang, Niranjan, Nelson, Zhang, Su, Folkman, Baek et~al.}}]{Jang-Science-2012}
\bibinfo{author}{\bibfnamefont{H.~W.} \bibnamefont{Jang}},
  \bibinfo{author}{\bibfnamefont{D.~A.} \bibnamefont{Felker}},
  \bibinfo{author}{\bibfnamefont{C.~W.} \bibnamefont{Bark}},
  \bibinfo{author}{\bibfnamefont{Y.}~\bibnamefont{Wang}},
  \bibinfo{author}{\bibfnamefont{M.~K.} \bibnamefont{Niranjan}},
  \bibinfo{author}{\bibfnamefont{C.~T.} \bibnamefont{Nelson}},
  \bibinfo{author}{\bibfnamefont{Y.}~\bibnamefont{Zhang}},
  \bibinfo{author}{\bibfnamefont{D.}~\bibnamefont{Su}},
  \bibinfo{author}{\bibfnamefont{C.~M.} \bibnamefont{Folkman}},
  \bibinfo{author}{\bibfnamefont{S.~H.} \bibnamefont{Baek}},
  \bibnamefont{et~al.}, \bibinfo{journal}{Science}
  \textbf{\bibinfo{volume}{331}}, \bibinfo{pages}{886} (\bibinfo{year}{2011}).

\bibitem[{\citenamefont{Boris et~al.}(2011)\citenamefont{Boris, Matiks,
  Benckiser, Frano, Popovich, Hinkov, Wochner, Castro-Colin, Detemple, Malik
  et~al.}}]{Boris-Science-2011}
\bibinfo{author}{\bibfnamefont{A.~V.} \bibnamefont{Boris}},
  \bibinfo{author}{\bibfnamefont{Y.}~\bibnamefont{Matiks}},
  \bibinfo{author}{\bibfnamefont{E.}~\bibnamefont{Benckiser}},
  \bibinfo{author}{\bibfnamefont{A.}~\bibnamefont{Frano}},
  \bibinfo{author}{\bibfnamefont{P.}~\bibnamefont{Popovich}},
  \bibinfo{author}{\bibfnamefont{V.}~\bibnamefont{Hinkov}},
  \bibinfo{author}{\bibfnamefont{P.}~\bibnamefont{Wochner}},
  \bibinfo{author}{\bibfnamefont{M.}~\bibnamefont{Castro-Colin}},
  \bibinfo{author}{\bibfnamefont{E.}~\bibnamefont{Detemple}},
  \bibinfo{author}{\bibfnamefont{V.~K.} \bibnamefont{Malik}},
  \bibnamefont{et~al.}, \bibinfo{journal}{Science}
  \textbf{\bibinfo{volume}{332}}, \bibinfo{pages}{937} (\bibinfo{year}{2011}).

\bibitem[{\citenamefont{Brinkman et~al.}(2007)\citenamefont{Brinkman, Huijben,
  Zalk, Huijben, Zeitler, Maan, der Wiel, Rijnders, Blank, and
  Hilgenkamp}}]{Brinkman-NatMat-2007}
\bibinfo{author}{\bibfnamefont{A.}~\bibnamefont{Brinkman}},
  \bibinfo{author}{\bibfnamefont{M.}~\bibnamefont{Huijben}},
  \bibinfo{author}{\bibfnamefont{M.~V.} \bibnamefont{Zalk}},
  \bibinfo{author}{\bibfnamefont{J.}~\bibnamefont{Huijben}},
  \bibinfo{author}{\bibfnamefont{U.}~\bibnamefont{Zeitler}},
  \bibinfo{author}{\bibfnamefont{J.~C.} \bibnamefont{Maan}},
  \bibinfo{author}{\bibfnamefont{W.~G.~V.} \bibnamefont{der Wiel}},
  \bibinfo{author}{\bibfnamefont{G.}~\bibnamefont{Rijnders}},
  \bibinfo{author}{\bibfnamefont{D.~H.~A.} \bibnamefont{Blank}},
  \bibnamefont{and}
  \bibinfo{author}{\bibfnamefont{H.}~\bibnamefont{Hilgenkamp}},
  \bibinfo{journal}{Nature Mater.} \textbf{\bibinfo{volume}{6}},
  \bibinfo{pages}{493} (\bibinfo{year}{2007}).

\bibitem[{\citenamefont{Li et~al.}(2011)\citenamefont{Li, Richter, Mannhart,
  and Ashoori}}]{Li-NatPhys-2011}
\bibinfo{author}{\bibfnamefont{L.}~\bibnamefont{Li}},
  \bibinfo{author}{\bibfnamefont{C.}~\bibnamefont{Richter}},
  \bibinfo{author}{\bibfnamefont{J.}~\bibnamefont{Mannhart}}, \bibnamefont{and}
  \bibinfo{author}{\bibfnamefont{R.~C.} \bibnamefont{Ashoori}},
  \bibinfo{journal}{Nature Phys.} \textbf{\bibinfo{volume}{7}},
  \bibinfo{pages}{762} (\bibinfo{year}{2011}).

\bibitem[{\citenamefont{Bert et~al.}(2011)\citenamefont{Bert, Kalisky, Bell,
  Kim, Hikita, Hwang, and Moler}}]{Bert-NatPhys-2011}
\bibinfo{author}{\bibfnamefont{J.~A.} \bibnamefont{Bert}},
  \bibinfo{author}{\bibfnamefont{B.}~\bibnamefont{Kalisky}},
  \bibinfo{author}{\bibfnamefont{C.}~\bibnamefont{Bell}},
  \bibinfo{author}{\bibfnamefont{M.}~\bibnamefont{Kim}},
  \bibinfo{author}{\bibfnamefont{Y.}~\bibnamefont{Hikita}},
  \bibinfo{author}{\bibfnamefont{H.~Y.} \bibnamefont{Hwang}}, \bibnamefont{and}
  \bibinfo{author}{\bibfnamefont{K.~A.} \bibnamefont{Moler}},
  \bibinfo{journal}{Nature Phys.} \textbf{\bibinfo{volume}{7}},
  \bibinfo{pages}{767} (\bibinfo{year}{2011}).

\bibitem[{\citenamefont{Liu et~al.}(2011)\citenamefont{Liu, Okamoto, van
  Veenendaal, Kareev, Gray, Ryan, Freeland, and Chakhalian}}]{Liu-PRB-2011}
\bibinfo{author}{\bibfnamefont{J.}~\bibnamefont{Liu}},
  \bibinfo{author}{\bibfnamefont{S.}~\bibnamefont{Okamoto}},
  \bibinfo{author}{\bibfnamefont{M.}~\bibnamefont{van Veenendaal}},
  \bibinfo{author}{\bibfnamefont{M.}~\bibnamefont{Kareev}},
  \bibinfo{author}{\bibfnamefont{B.}~\bibnamefont{Gray}},
  \bibinfo{author}{\bibfnamefont{P.}~\bibnamefont{Ryan}},
  \bibinfo{author}{\bibfnamefont{J.~W.} \bibnamefont{Freeland}},
  \bibnamefont{and}
  \bibinfo{author}{\bibfnamefont{J.}~\bibnamefont{Chakhalian}},
  \bibinfo{journal}{Phys. Rev. B} \textbf{\bibinfo{volume}{83}},
  \bibinfo{pages}{161102} (\bibinfo{year}{2011}).

\bibitem[{\citenamefont{Liu et~al.}(2012)\citenamefont{Liu, Kareev, Meyers,
  Gray, Ryan, Freeland, and Chakhalian}}]{Liu-PRL-2012}
\bibinfo{author}{\bibfnamefont{J.}~\bibnamefont{Liu}},
  \bibinfo{author}{\bibfnamefont{M.}~\bibnamefont{Kareev}},
  \bibinfo{author}{\bibfnamefont{D.}~\bibnamefont{Meyers}},
  \bibinfo{author}{\bibfnamefont{B.}~\bibnamefont{Gray}},
  \bibinfo{author}{\bibfnamefont{P.}~\bibnamefont{Ryan}},
  \bibinfo{author}{\bibfnamefont{J.~W.} \bibnamefont{Freeland}},
  \bibnamefont{and}
  \bibinfo{author}{\bibfnamefont{J.}~\bibnamefont{Chakhalian}},
  \bibinfo{journal}{Phys. Rev. Lett.} \textbf{\bibinfo{volume}{109}},
  \bibinfo{pages}{107402} (\bibinfo{year}{2012}).

\bibitem[{\citenamefont{Monkman et~al.}(2012)\citenamefont{Monkman, Adamo,
  Mundy, Shai, Harter, Shen, Burganov, Muller, Schlom, and
  Shen}}]{Monkman-NatMat-2012}
\bibinfo{author}{\bibfnamefont{E.~J.} \bibnamefont{Monkman}},
  \bibinfo{author}{\bibfnamefont{C.}~\bibnamefont{Adamo}},
  \bibinfo{author}{\bibfnamefont{J.~A.} \bibnamefont{Mundy}},
  \bibinfo{author}{\bibfnamefont{D.~E.} \bibnamefont{Shai}},
  \bibinfo{author}{\bibfnamefont{J.~W.} \bibnamefont{Harter}},
  \bibinfo{author}{\bibfnamefont{D.}~\bibnamefont{Shen}},
  \bibinfo{author}{\bibfnamefont{B.}~\bibnamefont{Burganov}},
  \bibinfo{author}{\bibfnamefont{D.~A.} \bibnamefont{Muller}},
  \bibinfo{author}{\bibfnamefont{D.~G.} \bibnamefont{Schlom}},
  \bibnamefont{and} \bibinfo{author}{\bibfnamefont{K.~M.} \bibnamefont{Shen}},
  \bibinfo{journal}{Nature Mater.} \textbf{\bibinfo{volume}{11}},
  \bibinfo{pages}{855} (\bibinfo{year}{2012}).

\bibitem[{\citenamefont{Chakhalian et~al.}(2007)\citenamefont{Chakhalian,
  Freeland, Habermeier, Cristiani, Khaliullin, van Veenendaal, and
  Keimer}}]{Chakhalian-Science-2007}
\bibinfo{author}{\bibfnamefont{J.}~\bibnamefont{Chakhalian}},
  \bibinfo{author}{\bibfnamefont{J.~W.} \bibnamefont{Freeland}},
  \bibinfo{author}{\bibfnamefont{H.-U.} \bibnamefont{Habermeier}},
  \bibinfo{author}{\bibfnamefont{G.}~\bibnamefont{Cristiani}},
  \bibinfo{author}{\bibfnamefont{G.}~\bibnamefont{Khaliullin}},
  \bibinfo{author}{\bibfnamefont{M.}~\bibnamefont{van Veenendaal}},
  \bibnamefont{and} \bibinfo{author}{\bibfnamefont{B.}~\bibnamefont{Keimer}},
  \bibinfo{journal}{Science} \textbf{\bibinfo{volume}{318}},
  \bibinfo{pages}{1115} (\bibinfo{year}{2007}).

\bibitem[{\citenamefont{Chakhalian et~al.}(2006)\citenamefont{Chakhalian,
  Freeland, Srajer, Strempfer, Khaliullin, Cezar, Charlton, Dalgliesh,
  Bernhard, Cristiani et~al.}}]{Chakhalian-NatPhys-2006}
\bibinfo{author}{\bibfnamefont{J.}~\bibnamefont{Chakhalian}},
  \bibinfo{author}{\bibfnamefont{J.~W.} \bibnamefont{Freeland}},
  \bibinfo{author}{\bibfnamefont{G.}~\bibnamefont{Srajer}},
  \bibinfo{author}{\bibfnamefont{J.}~\bibnamefont{Strempfer}},
  \bibinfo{author}{\bibfnamefont{G.}~\bibnamefont{Khaliullin}},
  \bibinfo{author}{\bibfnamefont{J.~C.} \bibnamefont{Cezar}},
  \bibinfo{author}{\bibfnamefont{T.}~\bibnamefont{Charlton}},
  \bibinfo{author}{\bibfnamefont{R.}~\bibnamefont{Dalgliesh}},
  \bibinfo{author}{\bibfnamefont{C.}~\bibnamefont{Bernhard}},
  \bibinfo{author}{\bibfnamefont{G.}~\bibnamefont{Cristiani}},
  \bibnamefont{et~al.}, \bibinfo{journal}{Nature Phys.}
  \textbf{\bibinfo{volume}{2}}, \bibinfo{pages}{244} (\bibinfo{year}{2006}).

\bibitem[{\citenamefont{Garcia-Barriocanal
  et~al.}(2010)\citenamefont{Garcia-Barriocanal, Cezar, Bruno, Thakur, Brookes,
  Utfeld, Rivera-Calzada, Giblin, Taylor, Duffy et~al.}}]{Garcia-NatComm-2010}
\bibinfo{author}{\bibfnamefont{J.}~\bibnamefont{Garcia-Barriocanal}},
  \bibinfo{author}{\bibfnamefont{J.}~\bibnamefont{Cezar}},
  \bibinfo{author}{\bibfnamefont{F.}~\bibnamefont{Bruno}},
  \bibinfo{author}{\bibfnamefont{P.}~\bibnamefont{Thakur}},
  \bibinfo{author}{\bibfnamefont{N.}~\bibnamefont{Brookes}},
  \bibinfo{author}{\bibfnamefont{C.}~\bibnamefont{Utfeld}},
  \bibinfo{author}{\bibfnamefont{A.}~\bibnamefont{Rivera-Calzada}},
  \bibinfo{author}{\bibfnamefont{S.}~\bibnamefont{Giblin}},
  \bibinfo{author}{\bibfnamefont{J.}~\bibnamefont{Taylor}},
  \bibinfo{author}{\bibfnamefont{J.}~\bibnamefont{Duffy}},
  \bibnamefont{et~al.}, \bibinfo{journal}{Nat. Commun.}
  \textbf{\bibinfo{volume}{1}}, \bibinfo{pages}{82} (\bibinfo{year}{2010}).

\bibitem[{\citenamefont{Gibert et~al.}(2012)\citenamefont{Gibert, Zubko1,
  Scherwitzl, Iniguez, and Triscone}}]{Gibert-NatMat-2012}
\bibinfo{author}{\bibfnamefont{M.}~\bibnamefont{Gibert}},
  \bibinfo{author}{\bibfnamefont{P.}~\bibnamefont{Zubko1}},
  \bibinfo{author}{\bibfnamefont{R.}~\bibnamefont{Scherwitzl}},
  \bibinfo{author}{\bibfnamefont{J.}~\bibnamefont{Iniguez}}, \bibnamefont{and}
  \bibinfo{author}{\bibfnamefont{J.-M.} \bibnamefont{Triscone}},
  \bibinfo{journal}{Nat. Mater.} \textbf{\bibinfo{volume}{22}},
  \bibinfo{pages}{195} (\bibinfo{year}{2012}).

\bibitem[{\citenamefont{Hoffman et~al.}(2013)\citenamefont{Hoffman, Tung,
  Nelson-Cheeseman, Liu, Freeland, and Bhattacharya}}]{Hoffman-2013}
\bibinfo{author}{\bibfnamefont{J.}~\bibnamefont{Hoffman}},
  \bibinfo{author}{\bibfnamefont{I.~C.} \bibnamefont{Tung}},
  \bibinfo{author}{\bibfnamefont{B.}~\bibnamefont{Nelson-Cheeseman}},
  \bibinfo{author}{\bibfnamefont{M.}~\bibnamefont{Liu}},
  \bibinfo{author}{\bibfnamefont{J.}~\bibnamefont{Freeland}}, \bibnamefont{and}
  \bibinfo{author}{\bibfnamefont{A.}~\bibnamefont{Bhattacharya}},
  \bibinfo{journal}{arXiv:1301.7295}  (\bibinfo{year}{2013}).

\bibitem[{\citenamefont{Lee and Han}(2013)}]{Lee-2013}
\bibinfo{author}{\bibfnamefont{A.~T.} \bibnamefont{Lee}} \bibnamefont{and}
  \bibinfo{author}{\bibfnamefont{M.~J.} \bibnamefont{Han}},
  \bibinfo{journal}{arXiv:1304.6555}  (\bibinfo{year}{2013}).

\bibitem[{\citenamefont{Serrate et~al.}(2007)\citenamefont{Serrate, Teresa, and
  Ibarra}}]{Serrate-JOP-2007}
\bibinfo{author}{\bibfnamefont{D.}~\bibnamefont{Serrate}},
  \bibinfo{author}{\bibfnamefont{J.~M.~D.} \bibnamefont{Teresa}},
  \bibnamefont{and} \bibinfo{author}{\bibfnamefont{M.~R.}
  \bibnamefont{Ibarra}}, \bibinfo{journal}{Journal of Physics: Condensed
  Matter} \textbf{\bibinfo{volume}{19}}, \bibinfo{pages}{023201}
  (\bibinfo{year}{2007}).

\bibitem[{\citenamefont{Rogado et~al.}(2005)\citenamefont{Rogado, Li, Sleight,
  and Subramanian}}]{Rogado-AdvMater-2005}
\bibinfo{author}{\bibfnamefont{N.}~\bibnamefont{Rogado}},
  \bibinfo{author}{\bibfnamefont{J.}~\bibnamefont{Li}},
  \bibinfo{author}{\bibfnamefont{A.}~\bibnamefont{Sleight}}, \bibnamefont{and}
  \bibinfo{author}{\bibfnamefont{M.}~\bibnamefont{Subramanian}},
  \bibinfo{journal}{Adv. Mater.} \textbf{\bibinfo{volume}{17}},
  \bibinfo{pages}{2225} (\bibinfo{year}{2005}).

\bibitem[{\citenamefont{Holmana et~al.}(2007)\citenamefont{Holmana, Huang,
  Klimczuk, Trzebiatowski, Bos, Morosan, Lynn, and Cavaa}}]{Holmana-JSSC-2007}
\bibinfo{author}{\bibfnamefont{K.}~\bibnamefont{Holmana}},
  \bibinfo{author}{\bibfnamefont{Q.}~\bibnamefont{Huang}},
  \bibinfo{author}{\bibfnamefont{T.}~\bibnamefont{Klimczuk}},
  \bibinfo{author}{\bibfnamefont{K.}~\bibnamefont{Trzebiatowski}},
  \bibinfo{author}{\bibfnamefont{J.}~\bibnamefont{Bos}},
  \bibinfo{author}{\bibfnamefont{E.}~\bibnamefont{Morosan}},
  \bibinfo{author}{\bibfnamefont{J.}~\bibnamefont{Lynn}}, \bibnamefont{and}
  \bibinfo{author}{\bibfnamefont{R.}~\bibnamefont{Cavaa}},
  \bibinfo{journal}{Journal of Solid Chemistry} \textbf{\bibinfo{volume}{180}},
  \bibinfo{pages}{75} (\bibinfo{year}{2007}).

\bibitem[{\citenamefont{Chaloupka and Khaliullin}(2008)}]{Chaloupka-PRL-2008}
\bibinfo{author}{\bibfnamefont{J.}~\bibnamefont{Chaloupka}} \bibnamefont{and}
  \bibinfo{author}{\bibfnamefont{G.}~\bibnamefont{Khaliullin}},
  \bibinfo{journal}{Phys. Rev. Lett.} \textbf{\bibinfo{volume}{100}},
  \bibinfo{pages}{016404} (\bibinfo{year}{2008}).

\bibitem[{\citenamefont{Hansmann et~al.}(2009)\citenamefont{Hansmann, Yang,
  Toschi, Khaliullin, Andersen, and Held}}]{Hansmann-PRL-2009}
\bibinfo{author}{\bibfnamefont{P.}~\bibnamefont{Hansmann}},
  \bibinfo{author}{\bibfnamefont{X.}~\bibnamefont{Yang}},
  \bibinfo{author}{\bibfnamefont{A.}~\bibnamefont{Toschi}},
  \bibinfo{author}{\bibfnamefont{G.}~\bibnamefont{Khaliullin}},
  \bibinfo{author}{\bibfnamefont{O.~K.} \bibnamefont{Andersen}},
  \bibnamefont{and} \bibinfo{author}{\bibfnamefont{K.}~\bibnamefont{Held}},
  \bibinfo{journal}{Phys. Rev. Lett.} \textbf{\bibinfo{volume}{103}},
  \bibinfo{pages}{016401} (\bibinfo{year}{2009}).

\bibitem[{\citenamefont{Mizokawa and Fujimori}(1995)}]{Mizokawa-PRB-1995}
\bibinfo{author}{\bibfnamefont{T.}~\bibnamefont{Mizokawa}} \bibnamefont{and}
  \bibinfo{author}{\bibfnamefont{A.}~\bibnamefont{Fujimori}},
  \bibinfo{journal}{Phys. Rev. B} \textbf{\bibinfo{volume}{51}},
  \bibinfo{pages}{12880} (\bibinfo{year}{1995}).

\bibitem[{\citenamefont{Okatov et~al.}(2005)\citenamefont{Okatov, Poteryaev,
  and Lichtenstein}}]{Okatov-EPL-2005}
\bibinfo{author}{\bibfnamefont{S.}~\bibnamefont{Okatov}},
  \bibinfo{author}{\bibfnamefont{A.}~\bibnamefont{Poteryaev}},
  \bibnamefont{and}
  \bibinfo{author}{\bibfnamefont{A.}~\bibnamefont{Lichtenstein}},
  \bibinfo{journal}{Europhysics Letters} \textbf{\bibinfo{volume}{70}},
  \bibinfo{pages}{499} (\bibinfo{year}{2005}).

\bibitem[{\citenamefont{Marianetti et~al.}(2004)\citenamefont{Marianetti,
  Kotliar, and Ceder}}]{Marianetti-PRL-2004}
\bibinfo{author}{\bibfnamefont{C.~A.} \bibnamefont{Marianetti}},
  \bibinfo{author}{\bibfnamefont{G.}~\bibnamefont{Kotliar}}, \bibnamefont{and}
  \bibinfo{author}{\bibfnamefont{G.}~\bibnamefont{Ceder}},
  \bibinfo{journal}{Phys. Rev. Lett.} \textbf{\bibinfo{volume}{92}},
  \bibinfo{pages}{196405} (\bibinfo{year}{2004}).

\bibitem[{\citenamefont{Torrance et~al.}(1992)\citenamefont{Torrance, Lacorre,
  Nazzal, Ansaldo, and Niedermayer}}]{Torrance-PRB-1992}
\bibinfo{author}{\bibfnamefont{J.~B.} \bibnamefont{Torrance}},
  \bibinfo{author}{\bibfnamefont{P.}~\bibnamefont{Lacorre}},
  \bibinfo{author}{\bibfnamefont{A.~I.} \bibnamefont{Nazzal}},
  \bibinfo{author}{\bibfnamefont{E.~J.} \bibnamefont{Ansaldo}},
  \bibnamefont{and}
  \bibinfo{author}{\bibfnamefont{C.}~\bibnamefont{Niedermayer}},
  \bibinfo{journal}{Phys. Rev. B} \textbf{\bibinfo{volume}{45}},
  \bibinfo{pages}{8209} (\bibinfo{year}{1992}).

\bibitem[{\citenamefont{Gou et~al.}(2011)\citenamefont{Gou, Grinberg, Rappe,
  and Rondinelli}}]{Gou-PRB-2011}
\bibinfo{author}{\bibfnamefont{G.}~\bibnamefont{Gou}},
  \bibinfo{author}{\bibfnamefont{I.}~\bibnamefont{Grinberg}},
  \bibinfo{author}{\bibfnamefont{A.~M.} \bibnamefont{Rappe}}, \bibnamefont{and}
  \bibinfo{author}{\bibfnamefont{J.~M.} \bibnamefont{Rondinelli}},
  \bibinfo{journal}{Phys. Rev. B} \textbf{\bibinfo{volume}{84}},
  \bibinfo{pages}{144101} (\bibinfo{year}{2011}).

\bibitem[{\citenamefont{Benckiser et~al.}(2011)\citenamefont{Benckiser,
  Haverkort, Brück, Goering, Macke, Frañó, Yang, Andersen, Cristiani,
  Habermeier et~al.}}]{Benckiser-NatMater-2011}
\bibinfo{author}{\bibfnamefont{E.}~\bibnamefont{Benckiser}},
  \bibinfo{author}{\bibfnamefont{M.~W.} \bibnamefont{Haverkort}},
  \bibinfo{author}{\bibfnamefont{S.}~\bibnamefont{Brück}},
  \bibinfo{author}{\bibfnamefont{E.}~\bibnamefont{Goering}},
  \bibinfo{author}{\bibfnamefont{S.}~\bibnamefont{Macke}},
  \bibinfo{author}{\bibfnamefont{A.}~\bibnamefont{Frañó}},
  \bibinfo{author}{\bibfnamefont{X.}~\bibnamefont{Yang}},
  \bibinfo{author}{\bibfnamefont{O.~K.} \bibnamefont{Andersen}},
  \bibinfo{author}{\bibfnamefont{G.}~\bibnamefont{Cristiani}},
  \bibinfo{author}{\bibfnamefont{H.-U.} \bibnamefont{Habermeier}},
  \bibnamefont{et~al.}, \bibinfo{journal}{Nat. Mater.}
  \textbf{\bibinfo{volume}{10}}, \bibinfo{pages}{189} (\bibinfo{year}{2011}).

\bibitem[{\citenamefont{Han et~al.}(2011)\citenamefont{Han, Wang, Marianetti,
  and Millis}}]{Han-PRL-2011}
\bibinfo{author}{\bibfnamefont{M.~J.} \bibnamefont{Han}},
  \bibinfo{author}{\bibfnamefont{X.}~\bibnamefont{Wang}},
  \bibinfo{author}{\bibfnamefont{C.~A.} \bibnamefont{Marianetti}},
  \bibnamefont{and} \bibinfo{author}{\bibfnamefont{A.~J.}
  \bibnamefont{Millis}}, \bibinfo{journal}{Phys. Rev. Lett.}
  \textbf{\bibinfo{volume}{107}}, \bibinfo{pages}{206804}
  (\bibinfo{year}{2011}).

\bibitem[{\citenamefont{Anisimov et~al.}(2002)\citenamefont{Anisimov, Nekrasov,
  Kondakov, Rice, and Sigrist}}]{Anisimov-EPJB-2002}
\bibinfo{author}{\bibfnamefont{V.~I.} \bibnamefont{Anisimov}},
  \bibinfo{author}{\bibfnamefont{I.~A.} \bibnamefont{Nekrasov}},
  \bibinfo{author}{\bibfnamefont{D.~E.} \bibnamefont{Kondakov}},
  \bibinfo{author}{\bibfnamefont{T.~M.} \bibnamefont{Rice}}, \bibnamefont{and}
  \bibinfo{author}{\bibfnamefont{M.}~\bibnamefont{Sigrist}},
  \bibinfo{journal}{Eur. Phys. J. B} \textbf{\bibinfo{volume}{25}},
  \bibinfo{pages}{192} (\bibinfo{year}{2002}).

\bibitem[{\citenamefont{de' Medici et~al.}(2009)\citenamefont{de' Medici,
  Hassan, Capone, and Dai}}]{DeMedici-PRL-2009}
\bibinfo{author}{\bibfnamefont{L.}~\bibnamefont{de' Medici}},
  \bibinfo{author}{\bibfnamefont{S.~R.} \bibnamefont{Hassan}},
  \bibinfo{author}{\bibfnamefont{M.}~\bibnamefont{Capone}}, \bibnamefont{and}
  \bibinfo{author}{\bibfnamefont{X.}~\bibnamefont{Dai}},
  \bibinfo{journal}{Phys. Rev. Lett.} \textbf{\bibinfo{volume}{102}},
  \bibinfo{pages}{126401} (\bibinfo{year}{2009}).

\end{thebibliography}

\end{document}